# Platooning as a Service (PlaaS): A Sustainable Transportation Framework for Connected and Autonomous Vehicles

Bhosale Akshay Tanaji[a], Sayak Roychowdhury[b], Anand Abraham[b]
[a] Department of Information Systems & Business Analytics, Indian Institute of Management, Amritsar, Punjab, India
[b] Department of Industrial and Systems Engineering, Indian Institute of Technology, Kharagpur, West Bengal, India,
(Email: akshayb@iimamritsar.ac.in, sroychowdhury@iem.iitkgp.ac.in, anandabraham@iem.iitkgp.ac.in)

## Abstract

Vehicular Platooning is a well-known transportation technique that helps reduce fuel consumption, carbon emissions, and road congestion. When integrated with Connected and autonomous vehicle (CAV) technologies, platooning enhances the overall safety and efficiency of the transportation system. This article presents Platooning as a Service (PlaaS) platform, a decision-support framework to promote sustainable transportation through platooning. We have formulated this problem as a Stackelberg game, with the platoon service provider (PSP) as the leader and the service users as the followers. The PSP sets the pricing policy, and the follower responds by choosing the distance to be travelled with the platoon. The optimal service contract between the PSP and the follower vehicle is established with the Stackelberg equilibrium, which is derived using Karush-Kuhn-Tucker optimality conditions. Additionally, we have examined the impact of government subsidies on the PlaaS platform in reducing carbon emissions. Our model has been applied to a specific example problem to illustrate the benefits for both players. We have also derived managerial insights through sensitivity analysis, exploring the effect of different velocity levels, vehicle dimensions, government subsidies, and operational urgency on the players' utility and $CO_2$ emissions. Our analysis shows that PSP gains a higher profit from high delay-cost vehicles performing time-critical operations with higher platoon velocity. However, the benefits related to fuel consumption are only realized at moderate platoon velocities.

## 1. Introduction

### *1.1 Motivation*

Traffic intensity is escalating in most regions globally, resulting in a growing problem of traffic congestion. This trend coincides with a rise in demand for transportation services, resulting in higher fossil fuel consumption, emissions of harmful exhaust gases, and more complex traffic patterns. Road accidents already cause 1.3 million deaths annually, a figure projected to reach 1.9 million by 2030 [1]. Simultaneously, nations are committed to reducing greenhouse gas emissions as per the Kyoto Protocol. In the transportation sector, labor cost accounts for 38% of total operational costs, while fuel expenses make up about 30% [2]. Autonomous vehicle platooning offers a promising solution for reducing road accidents, labor cost, fuel consumption, and greenhouse gas emissions. Projects like SCANIA, PATH, SARTRE, and KONVOI study the effect of platooning on fuel consumption, $CO_2$ emission, and transportation capacity [3], [4].

An autonomous vehicle platoon involves a fleet of virtually linked vehicles traveling with low headway, facilitated by connected and automated vehicle (CAV) technologies like cooperative adaptive cruise control (CACC) and reliable vehicle-to-vehicle (V2V) communication [5]. Generally, an autonomous platoon configuration consists of a platoon leader (PL) and platoon followers (PF), which are separated by inter-vehicle distance (IVD). Two platoons are the inter-platoon distance (IPD) away from each other, as shown in Fig. 1. This coordination in driving enhances safety [6], [7], fuel efficiency [8], [9], labor utilization [9], [10], and traffic flow [11] in CAVs. This additional benefit of platooning increases the interest of different stakeholders in providing a platooning as a service (PlaaS) framework. In this setup, we assume that the leader vehicle is owned by a platoon service provider (PSP) and driven by a human driver, whereas follower vehicles are semi-autonomous. Despite its technical feasibility, the adoption of autonomous truck platooning is progressing slowly due to the involvement of multiple stakeholders, vehicle types, and the absence of policy-driven initiatives.

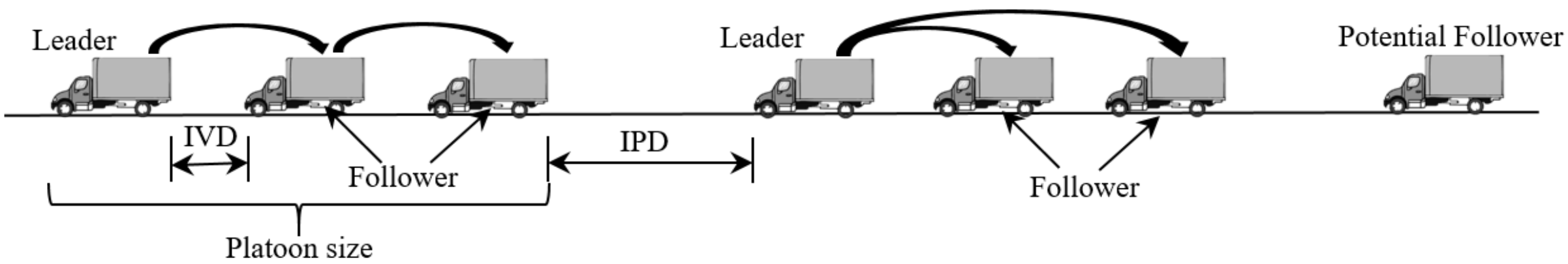

**Fig 1.** Terminology used in the platoon

From a practical standpoint, for successful platoon formation, both PSP and FV must be properly incentivized to make platooning operations acceptable to the stakeholders. The PSP aims to create a safe and secure platoon system to increase platoon formation and improve the benefits [7], [12]. In contrast, the follower vehicles attempt to minimize travel cost and reach the destination without delay. For the government, platooning offers an efficient solution towards higher road capacity utilization and a reduction in carbon emission. In connection with this, the following research questions are formulated:

RQ 1) How to quantify the benefits of autonomous platooning and develop a viable service model from the same?

RQ 2) What will be the optimal service contract between the platoon service provider and a vehicle joining the platoon under the assumption of perfect information?

RQ 3) How can government incentive policies influence platooning operations to develop sustainable transportation solutions?

This article aims to address these research questions by proposing the PlaaS framework, which considers the operational costs and benefits for the PSP and the follower vehicles. In this article, we have modeled the interaction between the PSP and a vehicle considering to join the platoon as a follower as a Stackelberg game. Both players act strategically to maximize their respective utilities. While the phrase "platooning as a service (PlaaS)" is already available in the literature [13], our proposed framework, PlaaS, delves into the economic viability for the first time. From now onwards, we will use the acronym FV to refer to potential follower vehicles, or vehicles considering to join the platoon as followers.

### *1.2 Contribution*

The major contributions of this article are summarized as follows:

- Develop CAV platooning as a service (PlaaS) framework involving the platoon leader as the representative of the PSP and the FVs as the potential subscribers of the service.
- Develop optimal cooperative policies for both PSP and the FV, with the PSP announces a service fee to the FV, and the FV calculate the distance to be travelled with the platoon.

- Analysis of the effectiveness of government subsidies as a mechanism for reducing carbon emissions in road transportation through the CAV PlaaS framework. Derive managerial insights through sensitivity analysis.

The remainder of the article is organized in the following way. Section 2 provides the literature review related to the platoon formation and characteristics of the platoon service. Section 3 explains the notations, assumptions, and formulates the problem for PSP and follower vehicles. Section 4 presents the proposed methodology and equilibrium strategies for the Stackelberg game. Section 5 provides numerical illustration and results for the sensitivity analysis. Section 6 highlights the managerial implications of the platooning service in the intelligent transportation system. Finally, the conclusion and future scope of this study are presented in Section 7.

## 2. Literature Review

In literature related to platooning, a substantial body of work is available on platoon scheduling, inter-vehicle and inter-platoon distance, and platoon formation to achieve stability and management. Review articles on platooning explore the factors influencing platoon formation, how different vehicle types and positions affect energy consumption, and the implications of autonomous platooning scenarios [6].

Lioris et al. [14] analyze the saturation flow capacity at an intersection and prove that platooning increases the intersection capacity by two or three times. Scholl et al. [15] propose an optimal platoon formation strategy that handles charging and platoon formation schedules for electric commercial vehicles. Zhou and Zhu [16] show that while larger platoon sizes increase road capacity, they reduce traffic flow stability; an optimal balance is achieved with platoons of ten or fewer vehicles. The game theory techniques address the trade-off between the cost sharing and investigate the factors essential for platoon formation [17], [18]. Sun and Yin [8] develop an optimal platoon formation strategy with an incentive mechanism to ensure platoon and vehicles' behavioral stability and willingness to join at different positions. Gattami et al. [19] apply optimal control and game theory to calculate the minimum safe relative distance needed in a platoon, balancing the trade-off between safety and fuel efficiency. Amoozadeh et al. [20] develop a platoon management protocol for platoon split, merge, and lane change operations. The protocol practices a centralized platoon coordination approach to ensure traffic flow stability and safe operation of the platoon. Ying et al. [21] propose a blockchain-based reputation management system using a multi-weighted subjective logic model for leader election, along with an incentive mechanism to reward high-reputation vehicles and promote active participation. Marzona et al. [22] analyse Italy's freight market and find that highly automated truck platooning is competitive with rail and conventional road transport for medium to long distances, highlighting its potential as an intelligent freight mode with notable market share. Lupi et al. [23] propose a two-echelon last-mile freight delivery system using platoons of automated electric vehicles, optimized through genetic algorithms and micro-simulation, showing reductions in delivery trips, congestion, emissions, and costs in the Lucca case study.

Xie et al. [24] propose a non-cooperative Stackelberg game among the leader vehicle (LV) and follower vehicles (FVs) to address the strategic interaction for incentive schemes. The incentive mechanism is based on the amount of data transmitted to LV and is calculated using a deep deterministic policy gradient (DDPG) algorithm. Yang et al. [25] investigate the computing and communication challenges of supporting vehicle platooning in an edge environment. They propose a contract-based incentive mechanism that encourages platoon formation and ensures optimal resource allocation between vehicles and the base station. Chen et al. [26] explore a cooperative platooning game to effectively distribute cost and profit among autonomous trucks by the exact row generation method. Lee et al. [27] propose an optimization model for heterogeneous platoons, showing that a bell-shaped formation maximizes energy savings and reduces $CO_2$ emissions. Pratibha et al. [28] propose a deep reinforcement learning (DRL) and genetic algorithm (GA) for achieving smart platooning (e.g. IVD, IPD) considering computational load and dynamic environment. Liatsos et al. [29] propose a capacitated hybrid truck platooning network design problem focused on driver delay penalties, showing significant cost savings over shortest-path routing and identifying an optimal platoon size of 4–6 trucks. Jo et al. [30] propose a unified methodology to evaluate the effects of truck platooning on freeway capacity and travel times. Applied to Korean freeways, the method demonstrates capacity gains and annual travel time savings of up to 187.6 billion KRW (167.7 million USD), supporting policies for platooning implementation.

Li et al. [31] propose a trust-aware decentralized speed advisory system (TD-SAS) to minimize energy consumption in autonomous vehicle platoons. The system integrates blockchain-based trust management, a speed recommendation scheme, and a cooperative game-based incentive mechanism to optimize overall energy consumption. Chen et al. [32] develop a qualitative business model for an autonomous vehicle platoon service platform, considering the AD device industry, IoT device industry, and platoon service as key stakeholders. The study related to CAV platoon robust control strategies, string stability and platoon merge formation is demonstrated in the [33], [34]. Yang and Murase [35] propose a robust, sensor-based vehicle platoon control method with a unified spacing policy, ensuring stable spacing and string stability while minimizing inter-vehicle communication in platoon system. Bai et al. [36] study the effect of government subsidies on third-party platoon coordination service and pricing rule in truck platooning. Scherr et al. [37] develop a two-stage stochastic model for urban platooning networks, optimizing fleet size, scheduling, routing, and pricing under demand uncertainty to improve service provider profitability and customer savings. Studies have been conducted on mixed and normal traffic platoons to demonstrate their effectiveness in fuel and $CO_2$ emission reduction, cooperative switching, and string stability within platoon systems [38], [39].

The literature shows that platooning offers significant benefits, including reduced fuel consumption, lower emissions, and reduced cognitive load. As a result, PlaaS is expected to become a key solution for efficient and safe transportation in the future. The decision of a platoon service involves a trade-off between service charges and travel distance for potential follower vehicles.

### *2.1. Research Gaps*

From the literature discussed above, the following research gaps are identified:

- The existing literature on platoon formation primarily focuses on the optimal sequence of FVs, fuel efficiency models, platoon scheduling, and PL selection. However, considering operational costs and benefits, there is a need to study the cooperative interaction between the PSP and FVs.
- One of the gaps observed in the existing literature is the lack of models that thoroughly consider the individual cost elements in a platoon system. So, there is a significant need to develop quantitative cost and benefit models informed by real-world studies and reports.
- There is a scarcity of studies on the impact of government subsidies on $CO_2$ emission reduction in platoon systems. Therefore, it is essential to evaluate the environmental sustainability of platoon systems and the role of government policies in promoting sustainable transportation.

The following section elaborates on the proposed problem formulation and game theory methodology to derive an optimal policy for the PlaaS platform.

## 3. Problem Formulation

We consider a platoon, where the leader is a human-driven vehicle (HDV), and the FVs are semi-autonomous. We model the decision problem of the FV and PSP as a Stackelberg game with perfect information, in which both the PSP and the FV maximize their respective expected utility functions. The mathematical model for different cost components and the Stackelberg game between PSP-FV is presented below. The notation used for the PlaaS platform is explained in Table 1.

**Table 1.**
Notation used for the PlaaS platform framework

| Notation | Description |
|---|---|
| $c_d$ | Delay cost in ₹/hr |
| $c_f$ | Fuel cost in ₹/lit |
| $c_o$ | Cognitive load cost in ₹/hr$^2$ |
| $c_c$ | Computational load cost in ₹/TB$^2$ |
| $\gamma_f$ | Government subsidy given for follower in ₹/km |
| $\gamma_l$ | Government subsidy given for PSP in ₹/km |
| $v$ | Individual vehicle velocity when travelling alone in km/hr |
| $v_p$ | Velocity of follower vehicle in platoon in km/hr |
| $\beta$ | Ratio of $v_p$ to $v$ (velocity reduction factor for follower vehicle when travelling in platoon, $0 \leq \beta \leq 1$) |
| $D$ | Total distance individual vehicle will travel in km |
| $L_T$ | Total computational load for platoon in TB/km |
| $\xi$ | Fraction of computational load shared by platoon to follower vehicle |
| $\psi$ | Specific fuel consumption in kg/kWh |
| $\eta$ | Vehicle efficiency in km/L |
| $C_{df}$ | Coefficient of air drag for vehicle when travelling alone |
| $C_{dp}$ | Coefficient of air drag for follower vehicle when travelling in platoon |
| $\alpha$ | Ratio of $C_{dp}$ to $C_{df}$, reduction factor in air drag due to platooning |
| $A$ | Cross-sectional area of vehicle in m$^2$ |
| $\rho_{air}$ | Air density in kg/m$^3$ |
| $\rho_{diesel}$ | Diesel density in g/m$^3$ |
| **Variables** | |
| $d$ | Distance travelled by the follower vehicle with platoon in km |
| $c_p$ | Service fee collected by PSP from FV in ₹/km |

### *3.1. Cost Modelling*

This section covers the components that will serve as a basis for the expected travelling cost and profit for the follower vehicle and platoon leader (PSP). We assume every hour spent on the road is associated with a delay cost of ₹ $c_d$. Typically, the average velocity of a vehicle travelling independently is more than that when travelling with a platoon; hence, there will be additional delay cost if a vehicle travels with a platoon for the same distance. Fuel consumption is a function of the air density and cross-section (front) area and vehicle efficiency, which are combined to a constant $T$ given as below:

$$T = \frac{0.5\rho_{air}A\psi}{3.6^2\eta 1000\rho_{diesel}}$$

Fuel consumption is also dependent on the coefficient of air drag$(C_{df}, C_{dp})$, and the velocity of the vehicle. The total fuel cost is proportional to the distance travelled $d$ and to the square of the velocity $v$, as shown in Table 2 [40].

Cognitive load refers to the driver's fatigue due to prolonged driving tasks. This cost is incorporated into a cost function, which is proportional to the square of the time of manual driving [41]. Since time is given by $\frac{d}{v}$, the expression for cognitive load cost at $c_o$ ₹/hr$^2$ as shown in Table 2. Autonomous driving requires computational resources, which may be shared among the PSP and the FV as required. The computational cost is assumed to be proportional to the square of the computational load $L_T$ TB and the fraction $\xi$ of load shared by the FV and PSP [25]. The total computational cost is proportional to the distance travelled in autonomous mode in the platoon at a rate $c_c$ ₹/TB$^2$ as shown in Table 2.

**Table 2.**
Cost components related to potential FV and PSP

| Cost Component | Travelling Alone | Travelling with Platoon |
|---|---|---|
| Delay cost for FV | $\frac{(D-d)}{v} c_d$ | $\frac{d}{v_p} c_d = \frac{d}{\beta v} c_d$ |
| Fuel cost for FV | $TC_{df} v^2 (D-d) c_f$ | $TC_{dp} v_p^2 d c_f$ $= T\alpha C_{df} \beta^2 v^2 d c_f$ |
| Cognitive load cost for FV | $c_o \frac{(D-d)^2}{v^2}$ | NA |
| Computational load cost for FV | NA | $\frac{1}{2} c_c (\xi L_T)^2 d$ |
| Computational load cost for PSP | NA | $\frac{1}{2} c_c \left((1-\xi) L_T\right)^2 d$ |
| Service charge for FV | NA | $d c_p$ |
| Government subsidy for FV | NA | $\gamma_f$ |
| Government subsidy for PSP | NA | $\gamma_l$ |

Based on the different cost components defined in the previous section, we have developed a quadratic programming problem with distinct objectives for both players. While formulating the optimization problem, we have considered the following assumptions:

- The platoon is Homogeneous and dynamic.
- The platoon has complete information about vehicle type, velocity, fuel reduction factor, and total travelling distance of a vehicle.
- The cognitive load for an individual vehicle varies as a square function of travelling distance. Also, the cognitive load is reduced to zero after joining the platoon for a semi-autonomous follower vehicle.

*3.2. Follower vehicle's problem*

The semi-autonomous FVs operate as HDVs when traveling independently but function as CAVs within a platoon service. The platoon leader is responsible for all the driving operations and communication with the follower vehicle. In contrast, the semi-autonomous follower vehicle utilizes onboard sensors and computational systems to follow the lead vehicle. This approach facilitates a practical transition to full automation, capitalizing on existing infrastructure and driver experience while progressively incorporating autonomous functionalities.

Suppose the total distance to be travelled by the FV is $D\ km$, and it has to decide the distance that it should travel with the platoon i.e. $d\ km$ to minimize total travel cost, which is divided into two parts, the cost of travelling individually $(c_{ind})$ and the cost of travelling with the platoon $(c_{pla})$.

$$c_{ind} = \frac{(D-d)}{v} c_d + TC_{df} v^2 (D-d) c_f + c_o \frac{(D-d)^2}{v^2}$$

$$c_{pla} = \left(\frac{d}{v_p}\right) c_d + TC_{dp} v_p^2 d c_f + \frac{1}{2} c_c d (\xi L_T)^2 + d c_p - \gamma_f d$$

The FV's total travel cost is given by

$$c_{FV} = c_{ind} + c_{pla}$$

The FV's problem in the Stackelberg game is

$$\min_{d} c_{FV} \qquad .... (1)$$

Subject to: $\quad 0 \leq d \leq D$

The FV determines the optimal distance $d^*$ to be travelled with the platoon as the best response to the service fee $c_p$ set by the PSP, by solving the following optimization problem

$$d^{\#} = argmin_{\{d \in [0,D]\}} c_{FV}(d, c_p)$$

*3.3. Platoon service provider's (PSP) problem*

The PSP's objective is to maximize its profits. The PSP makes money through service fees from FV and government subsidies. The costs include the delay, computational, and cognitive load costs for the driver (equivalent to the driver's salary), and fuel costs. The profit of the PSP is given below.

$$w_{PSP} = c_p d + \gamma_l d - c_{d'} \left(\frac{d}{v_p}\right) - \frac{1}{2} c_c \big((1-\xi) L_T\big)^2 d - c_o \left(\frac{d^2}{v_p^2}\right) - TC_{df} v_p^2 d c_f$$

The PSP solves the following problem.

$$\max_{c_p} w_{PSP} \qquad .... (2)$$

Subject to, $\quad c_p \geq 0$

The PSP determines the optimal service fee $c_p^*$ to be levied from the FV, as the best response to the distance $d^*$ to be travelled by the FV, by solving the following optimization problem.

$$c_p^* = argmin_{\{c_p \geq 0\}} W_{PSP}(d^{\#}, c_p)$$

## 4. Solution methodology

Cooperative platoon formation allows both the follower vehicle and service provider to maximize their utility by selecting the traveling distance with a platoon $(d)$ for the follower vehicle and service fee $(c_P)$ for the leader. Accordingly, the PSP acts as the leader in this sequential decision-making process, first determines the per-kilometre platoon service charge $(c_p)$ to be imposed on the FV. In response to the announced charge, the FV subsequently decides on the distance it will travel as part of the platoon. Based on the optimal service charge $c_p^*$ of the PSP, the FV determines the optimal distance $d^*$. The sequence of steps is given in Fig. 2.

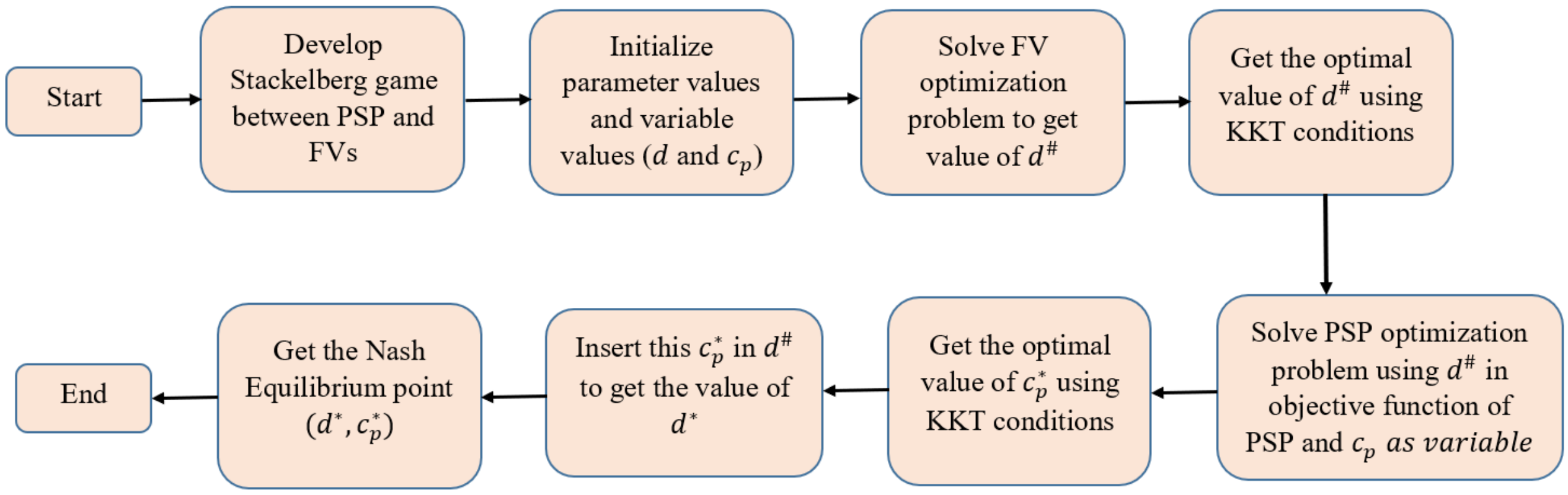

**Fig 2.** Flowchart of proposed methodology

*4.1 Equilibrium solution to the Stackelberg game*

**Proposition 1.1:** The FV's optimization problem given in Eq. (1) is a convex programming problem with respect to the decision variable $d$. (Proof of this is explained in Appendix A.1.1)
**Proposition 1.2:** For the follower vehicle's problem, there exists a global minimizer of travel distance $d^{\#}$, and Lagrangian multipliers $\lambda_1^*$ and $\lambda_2^*$, satisfying the following KKT conditions: (Proof of this is explained in Appendix A.1.2)

$$\begin{cases} -\dfrac{2c_o(D-d^{\#})}{v^2} - C_{df}Tc_f v^2 + C_{dp}Tc_f v_p^2 + \dfrac{c_d}{v_p} - \dfrac{c_d}{v} + c_p + \dfrac{1}{2}c_c(\xi L_T)^2 - \gamma_f - \lambda_1^* + \lambda_2^* = 0 \\ \lambda_1^*(-d^{\#}) = 0 \\ \lambda_2^*(d^{\#} - D) = 0 \\ -d^{\#} \leq 0 \\ d^{\#} - D \leq 0 \\ \lambda_1^*, \lambda_2^* \geq 0 \end{cases} \quad \dots(3)$$

**Remark 1**: The Lagrangian multipliers $\lambda_1$ and $\lambda_2$ are related to the minimum and maximum distance constraint in Eq. (1) respectively. These Lagrangian multipliers are associated with "shadow prices," i.e. the small increment ($\sigma$) in distance leads to a change in the optimal objective function by approximately $\lambda_1\sigma$ or $\lambda_2\sigma$ amount. As the given optimization problem is convex, all the KKT conditions are necessary and sufficient for the optimal solution. The equilibrium solutions presented in Proposition 1.2 represent the global optimal defense allocation for each target.

**Theorem 1:** The best response ($d^{\#}$) of the FV to the service fee $c_p$ set by the PSP is given by

$$d^{\#} = \begin{cases} 0 & if\ \nabla_d c_{FV}|_{d=0} > 0 \\ D & if\ \nabla_d c_{FV}|_{d=D} < 0 \\ D + \dfrac{v^2}{2c_0}\left(C_{AD}v^2(1-\alpha\beta^2) + \dfrac{c_d}{v}\left(1-\left(\dfrac{1}{\beta}\right)\right) - \dfrac{1}{2}c_c(\xi L_T)^2 + \gamma_f - c_p\right) & otherwise \end{cases}$$

Where $C_{AD} = TC_{df}c_f$ ; $\alpha = \frac{C_{dp}}{C_{df}}$; $\beta = \frac{v_p}{v}$;

**Proof.** From the KKT conditions in (3), since both $\lambda_1^*, \lambda_2^* \geq 0$, four cases are possible,
**Case 1:** $\lambda_1^* > 0, \lambda_2^* > 0$
This case is not possible, because if both $\lambda_1^* > 0, \lambda_2^* > 0$, then due to the complementary slackness conditions given in (3), the value of $d^*$ should be equal to 0, and $D$ at the same time, which is not possible.
**Case 2:** $\lambda_1^* > 0, \lambda_2^* = 0$

In this case, if $\lambda_1^* > 0$, due to the KKT condition $\lambda_1^*(-d^{\#}) = 0$, it implies that $d^{\#} = 0$
Implementing this into the derivative of the Lagrangian function in (3):

$$\nabla_d c_{FV}|_{d=0} - \lambda_1^* = 0$$
$$\rightarrow \; \lambda_1^* = \nabla_d c_{FV}|_{d=0} > 0$$

This proves that $d^{\#} = 0$, when $\nabla_d c_{FV}|_{d=0} > 0$.
**Case 3:** $\lambda_1^* = 0, \lambda_2^* > 0$
In this case, if $\lambda_2^* > 0$, due to the KKT condition $\lambda_2^*(d^{\#} - D) = 0$, it implies that $d^{\#} = D$
Implementing this into the derivative of the Lagrangian function in (3):

$$\nabla_d c_{FV}|_{d=D} + \lambda_2^* = 0$$
$$\rightarrow \; \lambda_2^* = -\nabla_d c_{FV}|_{d=D} > 0$$
$$\rightarrow \; \nabla_d c_{FV}|_{d=D} < 0$$

This proves that $d^{\#} = D$, when $\nabla_d c_{FV}|_{d=D} < 0$.
**Case 4:** $\lambda_1^* = 0, \lambda_2^* = 0$
In this case, from the KKT conditions and the derivative of the Lagrangian function in (3), we get:

$$-\frac{2c_o(D-d^{\#})}{v^2} - C_{df}Tc_f v^2 + C_{dp}Tc_f v_p^2 + \frac{c_d}{v_p} - \frac{c_d}{v} + c_p + \frac{1}{2}c_c(\xi L_T)^2 - \gamma_f = 0$$
$$\rightarrow d^{\#} = \frac{v^2}{2c_0}\left(Tc_f\left(C_{df}v^2 - C_{dp}v_p^2\right) + \frac{2c_0 D}{v^2} + c_d\left(\frac{1}{v} - \frac{1}{v_p}\right) - c_p - \frac{1}{2}c_c(\xi L_T)^2 + \gamma_f\right)$$
$$= D + \frac{v^2}{2c_0}\left(C_{AD}v^2(1-\alpha\beta^2) + \frac{c_d}{v}\left(1 - \left(\frac{1}{\beta}\right)\right) - \frac{1}{2}c_c(\xi L_T)^2 + \gamma_f - c_p\right)$$

This completes the proof of Theorem 1.
We can shorten the expression for $d^*$ in the following way:

$$d^{\#} = D + \frac{v^2}{2c_0}\left(G + \gamma_f - c_p\right) \ldots \quad (4)$$

Where $G = C_{AD}v^2(1-\alpha\beta^2) + \frac{c_d}{v}\left(1 - \left(\frac{1}{\beta}\right)\right) - \frac{1}{2}c_c(\xi L_T)^2$.

**Corollary 1.1:** The FV will not join the platoon ($d^{\#} = 0$) if the service fee set by the PSP is
$c_p > \frac{2c_0 D}{v^2} + C_{AD}v^2(1-\alpha\beta^2) + \frac{c_d}{v}\left(1 - \left(\frac{1}{\beta}\right)\right) - \frac{1}{2}c_c(\xi L_T)^2 + \gamma_f = \frac{2c_0 D}{v^2} + G + \gamma_f$
**Proof:** This follows from the first statement of Theorem 1.

$$d^{\#} = 0 \;\; if \; \nabla_d c_{FV}|_{d=0} > 0$$
$$\nabla_d c_{FV}|_{d=0} = -\frac{2c_o D}{v^2} - C_{df}Tc_f v^2 + C_{dp}Tc_f v_p^2 + \frac{c_d}{v_p} - \frac{c_d}{v} + c_p + \frac{1}{2}c_c(\xi L_T)^2 - \gamma_f > 0$$
$$\rightarrow \; c_p > \frac{2c_o D}{v^2} + Tc_f\left(C_{df}v^2 - C_{dp}v_p^2\right) + c_d\left(\frac{1}{v} - \frac{1}{v_p}\right) - \frac{1}{2}c_c(\xi L_T)^2 + \gamma_f$$
$$= \frac{2c_0 D}{v^2} + C_{AD}v^2(1-\alpha\beta^2) + \frac{c_d}{v}\left(1 - \left(\frac{1}{\beta}\right)\right) - \frac{1}{2}c_c(\xi L_T)^2 + \gamma_f$$
$$= \frac{2c_0 D}{v^2} + G + \gamma_f$$

**Corollary 1.2:** The FV will travel the entire distance $D$ with the platoon if the service fee set by the PSP is

$$c_p < \; C_{AD}v^2(1-\alpha\beta^2) + \frac{c_d}{v}\left(1 - \left(\frac{1}{\beta}\right)\right) - \frac{1}{2}c_c(\xi L_T)^2 + \gamma_f = G + \gamma_f$$

**Proof:** This follows from the second statement of Theorem 1.

$$d^{\#} = D \ \ if \ \nabla_d c_{FV}|_{d=D} < D$$

$$\nabla_d c_{FV}|_{d=D} = -C_{df} T c_f v^2 + C_{dp} T c_f v_p^2 + \frac{c_d}{v_p} - \frac{c_d}{v} + c_p + \frac{1}{2} c_c (\xi L_T)^2 - \gamma_f < 0$$

$$\rightarrow \ c_p < \ T c_f \left( C_{df} v^2 - C_{dp} v_p^2 \right) + c_d \left( \frac{1}{v} - \frac{1}{v_p} \right) - \frac{1}{2} c_c (\xi L_T)^2 + \gamma_f$$

$$= C_{AD} v^2 (1 - \alpha \beta^2) + \frac{c_d}{v} \left( 1 - \left( \frac{1}{\beta} \right) \right) - \frac{1}{2} c_c (\xi L_T)^2 + \gamma_f$$

$$= G + \gamma_f$$

**Corollary 1.3:** The government subsidy $\gamma_f$ for the FV has an upper-bound and a lower-bound to ensure feasibility of the FV's decision to travel atmost $D$distance with the platoon as given below:

$$c_p - G - \frac{2c_0 D}{v^2} \leq \gamma_f \leq c_p - G,$$

**Proof:** According to Theorem 1,

$$d^{\#} = \begin{cases} 0 & if \ \nabla_d c_{FV}|_{d=0} > 0 \\ D & if \ \nabla_d c_{FV}|_{d=D} < 0 \\ \frac{v^2}{2c_0} \left( C_{AD} v^2 (1 - \alpha \beta^2) + \frac{2c_0 D}{v^2} + \frac{c_d}{v} \left( 1 - \left( \frac{1}{\beta} \right) \right) - \frac{1}{2} c_c (\xi L_T)^2 + \gamma_f - c_p \right) & otherwise \end{cases}$$

This can be written as

$$d^{\#} = D + \frac{v^2}{2c_0} \left( C_{AD} v^2 (1 - \alpha \beta^2) + \frac{c_d}{v} \left( 1 - \left( \frac{1}{\beta} \right) \right) - \frac{1}{2} c_c (\xi L_T)^2 + \gamma_f - c_p \right)$$

$$= D + \frac{v^2}{2c_0} \left( G + \gamma_f - c_p \right) = D - \frac{v^2}{2c_0} \left( c_p - G - \gamma_f \right)$$

To satisfy $d^{\#} \geq 0$, $\quad \frac{v^2}{2c_0} \left( c_p - G - \gamma_f \right) \leq D$

$$\rightarrow \gamma_f \geq c_p - G - \frac{2c_0 D}{v^2}$$

To satisfy $d^{\#} \leq D$, $\quad \frac{v^2}{2c_0} \left( c_p - G - \gamma_F \right) \geq 0$

$$\rightarrow \gamma_f \leq c_p - G$$

Combining the above two inequalities

$$c_p - G - \frac{2c_0 D}{v^2} \leq \gamma_f \leq c_p - G$$

**Proposition 2.1:** The PSP optimization problem given in Eq. (2) is a convex optimization problem after substituting $d^{\#}$ in PSP's objective function. (Proof of this is explained in Appendix A.2.1).

**Proposition 2.2:** For the PSP problem, there exists a global maximizer $c_p^*$ and Lagrangian multiplier $\theta_1^*$ that satisfy the following KKT conditions: (Proof of this is explained in Appendix A.2.2).

$$\begin{cases} \nabla_{c_p} w_{PSP}(d^{\#})|_{c_p^*} - \theta_1^* = 0 \\ \theta_1^* \left( -c_p^* \right) = 0 \\ -c_p^* \leq 0 \\ \theta_1^* \geq 0 \end{cases} \quad \ldots\ldots.(5)$$

**Remark 2**: The Lagrangian multiplier $\theta_1$ correspond to the minimum service fee constraint in (2), respectively. The Lagrangian multiplier $\theta_1$ specify the shadow prices associated with the service fee of PSP, i.e. a small increment $(\delta)$ in PSP service fee leads to a change in the optimal objective function by approximately $\theta_1 \times \delta$ amount. As the given problem is a convex optimization problem, all the KKT conditions are necessary and sufficient for the optimal solution. The solutions presented in Proposition 2.2 are global optimal solutions for each attacker against each target.

**Theorem 2:** The service fee $c_p^*$ to be set by the PSP as the best response to the FV's optimal travel distance $d^\#$ is given by

$$c_p^* = \\ = \begin{cases} 0 & if\ \nabla_{c_p} w_{PSP}|_{c_p} > 0 \\ \frac{1}{\left(2+\frac{1}{\beta^2}\right)}\left[\left(G+\gamma_f\right)\left(1+\frac{1}{\beta^2}\right)+\frac{1}{2}c_c\left((1-\xi)L_T\right)^2+C_{AD}\beta^2v^2+\left(\frac{c_{d'}}{\beta v}\right)-\gamma_l+\frac{2c_0D}{v^2}\left(1+\frac{1}{\beta^2}\right)\right] & otherw \end{cases}$$

Where $d^\# = \frac{v^2}{2c_0}\left(G+\gamma_f-c_p\right)$, $G = C_{AD}v^2(1-\alpha\beta^2)+\frac{c_d}{v}\left(1-\left(\frac{1}{\beta}\right)\right)-\frac{1}{2}c_c(\xi L_T)^2$

**Proof:** The proof is similar to that of Theorem 1. From the KKT conditions given in (4), there will be two cases, viz. i) $\theta_1^* > 0$,ii) $\theta_1^* = 0$

For $\theta_1^* > 0$, we get $c_p^* = 0$ from the complementary slackness condition.

For $\theta_1^* = 0$, from the KKT conditions and the derivative of the Lagrangian function in (5), we get:

$$\nabla_{c_p} w_{PSP}\left(c_p^*, d^\#\right) = 0$$

$$\begin{aligned} \rightarrow -\frac{v^2c_p}{2c_0}+\left(\frac{v^2}{2c_0}\right)\left[-c_p-\frac{1}{2}c_c(\xi L_T)^2+\gamma_f+TC_{df}v_p^2c_f+c_d\left(1-\frac{1}{\beta}\right)\right] \\ +\frac{v^2}{v_p^2}\left(\left(\frac{v^2}{2c_0}\right)\left[-c_p-\frac{1}{2}c_c(\xi L_T)^2+\gamma_f+TC_{df}v_p^2c_f+c_d\left(1-\frac{1}{\beta}\right)\right]+D\right) \\ +\frac{1}{4c_0}c_c\left((1-\xi)L_T\right)^2-\frac{v^2\gamma_l}{2c_0}+\frac{TC_{df}v_p^2c_fv^2}{2c_0}+\frac{c_{d'}v^2}{2c_0v_p}+D=0 \\ \rightarrow\ c_p^* = \frac{1}{\left(2+\frac{1}{\beta^2}\right)}\left[\left(G+\gamma_f\right)\left(1+\frac{1}{\beta^2}\right)+\frac{1}{2}c_c\left((1-\xi)L_T\right)^2+C_{AD}\beta^2v^2+\left(\frac{c_{d'}}{\beta v}\right)-\gamma_l\right. \\ \left.+\frac{2c_0D}{v^2}\left(1+\frac{1}{\beta^2}\right)\right] \end{aligned}$$

This completes the proof of Theorem 2.

**Theorem 3:** Assuming the FV travels a non-zero distance with the platoon, there exists a Subgame Perfect Equilibrium $(c_p^*, d^*)$ for the Stackelberg game described in (1) and (2), in which

$$d^* = \frac{v^2}{2c_0}\left( G + \gamma_f - \frac{1}{\left(2 + \frac{1}{\beta^2}\right)}\left[\left(G + \gamma_f\right)\left(1 + \frac{1}{\beta^2}\right) + \frac{1}{2}c_c\left((1-\xi)L_T\right)^2 + C_{AD}\beta^2 v^2 + \left(\frac{c_{d'}}{\beta v}\right) - \gamma_l + \frac{2c_0 D}{v^2}\left(1 + \frac{1}{\beta^2}\right)\right]\right) + D$$

$$c_p^* = \frac{1}{\left(2 + \frac{1}{\beta^2}\right)}\left[\left(G + \gamma_f\right)\left(1 + \frac{1}{\beta^2}\right) + \frac{1}{2}c_c\left((1-\xi)L_T\right)^2 + C_{AD}\beta^2 v^2 + \left(\frac{c_{d'}}{\beta v}\right) - \gamma_l + \frac{2c_0 D}{v^2}\left(1 + \frac{1}{\beta^2}\right)\right]$$

**Proof.** In this setting, the PSP is the leader who sets the service fee $c_p$ and FV is the follower who decided on the distance $d$ to be travelled with the platoon.
Given the assumption of $d > 0$, we need to show the existence of a Nash Equilibrium for two cases, i) $d^* < D$, ii) $d^* = D$

**Case 1:** $d < D$

From Theorem 1 and Eq. (4), the FV's best response $d^*$ to the PSP's service fee $c_p$ is given by

$$d^{\#} = b_{FV}\left(c_p\right) = \frac{v^2}{2c_0}\left(G + \gamma_f - c_p\right) + D$$

From Theorem 2, the PSP's best response $c_p^*$ to the FV's travel distance $d^*$ is given by

$$c_p^* = b_{PSP}(d^*) = \frac{1}{\left(2 + \frac{1}{\beta^2}\right)}\left[\left(G + \gamma_f\right)\left(1 + \frac{1}{\beta^2}\right) + \frac{1}{2}c_c\left((1-\xi)L_T\right)^2 + C_{AD}\beta^2 v^2 + \left(\frac{c_{d'}}{\beta v}\right) - \gamma_l + \frac{2c_0 D}{v^2}\left(1 + \frac{1}{\beta^2}\right)\right]$$

$$c_p^* = b_{PSP}(d^*) = \frac{1}{\left(2 + \frac{1}{\beta^2}\right)}\left[\left(G + \gamma_f + \frac{2c_0 D}{v^2}\right)\left(1 + \frac{1}{\beta^2}\right) + \frac{1}{2}c_c\left((1-\xi)L_T\right)^2 + C_{AD}\beta^2 v^2 + \left(\frac{c_{d'}}{\beta v}\right) - \gamma_l\right]$$

Using backward induction, the FV's best response to PSP's service fee $c_p^*$ is given by

$$d^* = b_{FV}\left(c_p^*\right) = \frac{v^2}{2c_0}\left(G + \gamma_f - c_p^*\right) + D$$

$$= \frac{v^2}{2c_0}\left( G + \gamma_f - \frac{1}{\left(2 + \frac{1}{\beta^2}\right)}\left[\left(G + \gamma_f + \frac{2c_0 D}{v^2}\right)\left(1 + \frac{1}{\beta^2}\right) + \frac{1}{2}c_c\left((1-\xi)L_T\right)^2 + C_{AD}\beta^2 v^2 + \left(\frac{c_{d'}}{\beta v}\right) - \gamma_l\right]\right) + D \;\ldots (7)$$

Since $d^* = b_{FV}(c_p^*)$ and $c_p^* = b_{PSP}(d^*)$, the strategy profile $(d^*, c_p^*)$ is a Nash Equilibrium.

**Case 2:** $d^* = D$

From Corollary 1.2, $d^\# = D$, if $c_p < K$.

For this case, the PSP's problem becomes

$$\max_{c_p} w_{PSP}(D) \qquad (6)$$

Subject to,
$$c_p < K$$
$$c_p \geq 0$$

For this problem, there exists a global maximizer $c_p^*$ and Lagrangian multipliers $\mu_1^*$ and $\mu_2^*$ that satisfy the following KKT conditions:

$$\begin{cases} \nabla_{c_p} w_{PSP}(D)|_{c_p^*} - \mu_1^* + \mu_2^* = 0 \\ \mu_1^*(-c_p^*) = 0 \\ \mu_2^*(c_p^* - K) = 0 \\ -c_p^* \leq 0 \\ c_p^* - K \leq 0 \\ \mu_1^*, \mu_2^* \geq 0 \end{cases} \quad \ldots\ldots (7)$$

From the first KKT condition we get $D - \mu_1^* + \mu_2^* = 0$.

**Case 1:** $\mu_1^* = 0, \mu_2^* = 0$

This leads to $D = 0$, which we consider a trivial case as this only occurs when the FV has no distance to travel.

**Case 2:** $\mu_1^* = 0, \mu_2^* > 0$

This case is impossible as it would lead to $D + \mu_2^* = 0$, at $c_p = G + \gamma_f$

$D$ has to be negative as $\mu_2^* > 0$.

**Case 3:** $\mu_1^* > 0, \mu_2^* = 0$

This leads to $c_p = 0$, which is a trivial case.

**Case 4:** $\mu_1^* > 0, \mu_2^* > 0$

This is an impossible case as it would mean $c_p = 0$ and $c_p = G + \gamma_f$ at the same time.

*4.2 Impact of government subsidies policy on sustainable platoon service*

The FV and PSP problems discussed in the previous section consist of different cost and benefit components. In this section, we analyze the effect of each factor on the optimal value of distance $(d^*)$ and service fee $(c_p^*)$ by dividing into four subcases. The four subcases and results obtained after solving the KKT conditions are given in Table 3.

Fig. 3 and Fig. 4 provide better visualization of the impact of government subsidy of both PSP and FVs on the optimal platooning distance $d^*$ and the optimal service fee $c_p^*$. When no subsidy is provided, $d^*$ is at its least favourable levels due to the full cost utilization on follower vehicles. Subsidizing only the follower vehicles increases $d^*$ moderately with the magnitude of $\frac{v^2\gamma_f}{2c_0\left(2+\frac{1}{\beta^2}\right)}$ and allows PSPs to charge a higher $c_p^*$ with the increase of $\gamma_f\left(1+\frac{1}{\beta^2}\right)/\left(2+\frac{1}{\beta^2}\right)$, while subsidizing only the PSP reduces $c_p^*$ by $\gamma_l/\left(2+\frac{1}{\beta^2}\right)$ and increases $d^*$ further by $\frac{v^2}{2c_0}\left(\gamma_f+\frac{\gamma_l}{2+1/\beta^2}\right)$. The most significant improvement in both distance and service affordability occurs when both the follower vehicles and the PSP are given a government subsidy, maximizing $d^*$ and permitting a sustainable service provider fee. These results highlight the effectiveness of joint subsidies in enhancing participation and economic practicability of the platooning as a service.

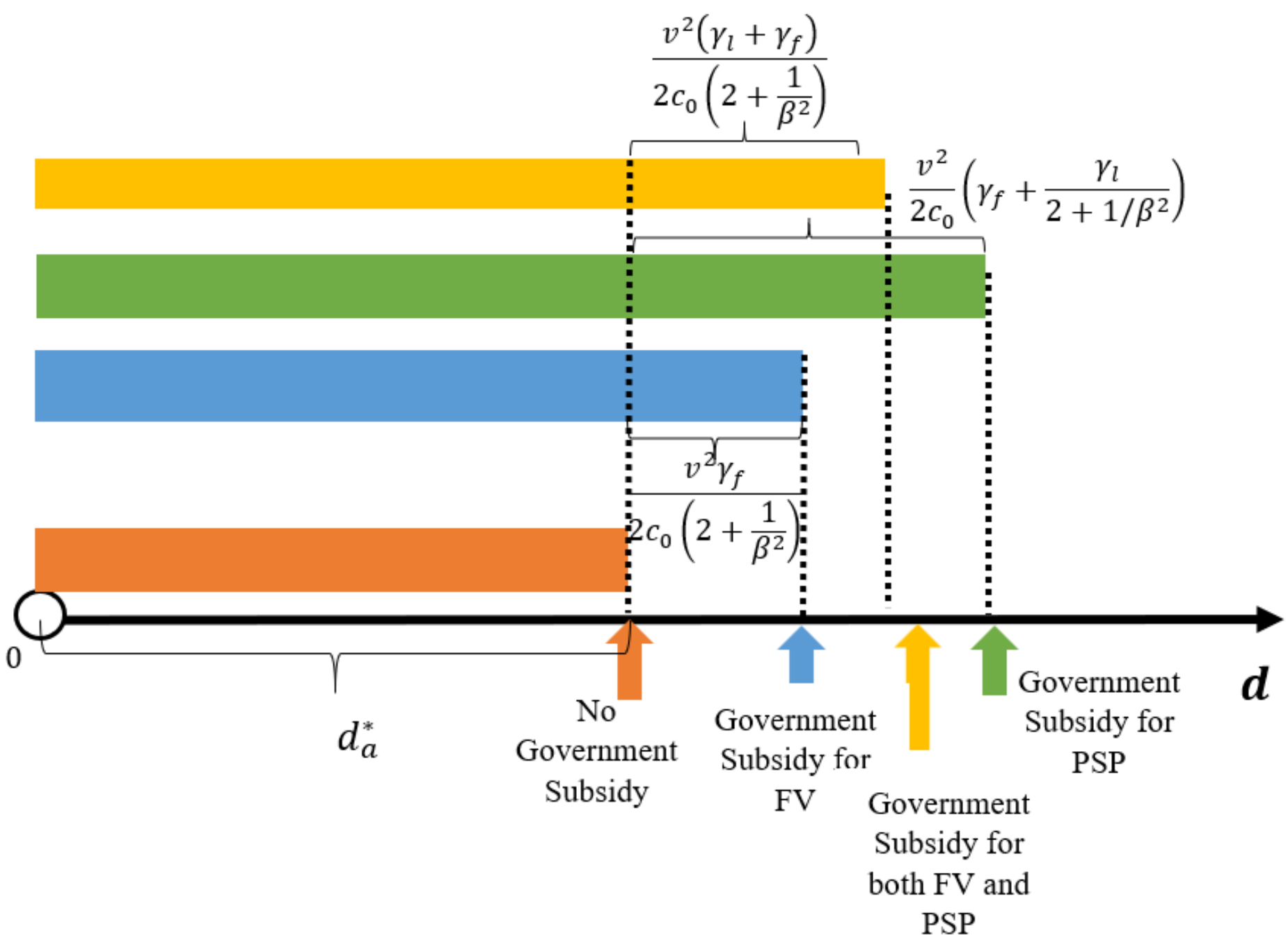

**Fig 3.** Effect of each cost and benefit on the optimal travel distance of FV with platoon $(d^*)$

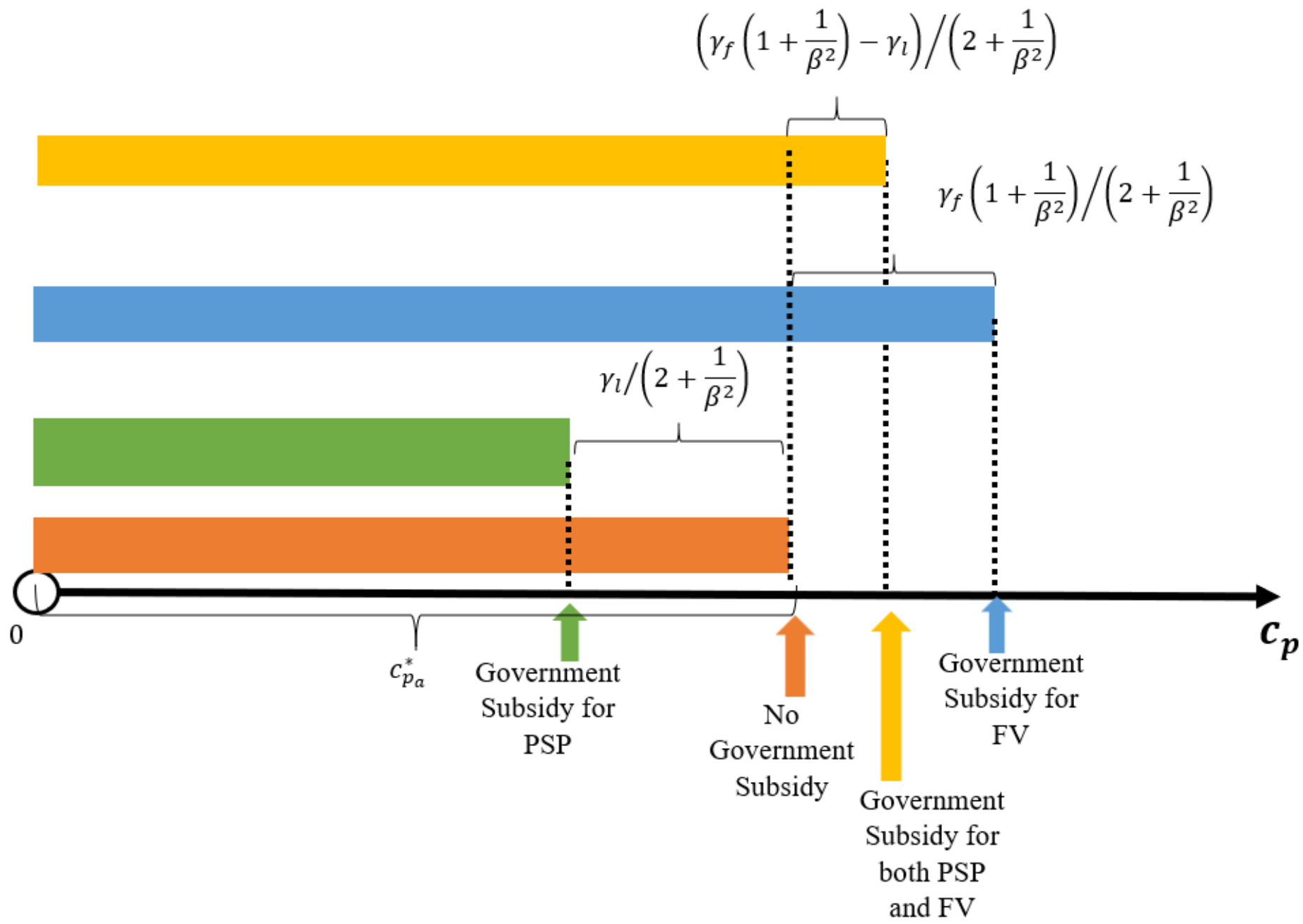

**Fig 4.** Effect of each cost and benefit on the optimal service fee of PSP $(c_p^*)$

**Proposition 3:** The introduction of government subsidies to both the leader and follower vehicles in a CAV platoon system leads to a reduction in $CO_2$ emission per trip, given by:

$$\Delta CO_2 = T\alpha\phi C_{df}\beta^2 v^2 \times \frac{v^2}{2c_0} \times \frac{\beta^2(\gamma_l+\gamma_f)}{(2\beta^2+1)}(1-\alpha\beta^2) \ldots\ldots (8)$$

where this saving is expressed in kilograms, assuming 1 litre of fuel consumption emits approximately $\phi$ kg of $CO_2$ (Proof of this is explained in Appendix A.3).

**Table 3.**
Equilibrium solutions for optimal travel distance and service fee for different platoon velocity

| **Case** | $\boldsymbol{c_p^*}$ | $\boldsymbol{d^*}$ |
|---|---|---|
| Without consideration of government subsidies for both players $(c_{pa}^*, d_a^*)$ | $max\left(0, \frac{1}{\left(2+\frac{1}{\beta^2}\right)}\left[\left(G+\frac{2c_0D}{v^2}\right)\left(1+\frac{1}{\beta^2}\right)+\frac{1}{2}c_c\left((1-\xi)L_T\right)^2+C_{AD}\beta^2v^2+\left(\frac{c_{d'}}{\beta v}\right)\right]\right)$ | $max\left(0, min\left(D, \frac{v^2}{2c_0}\left(G-c_p^*\right)+D\right)\right)$ |
| Considering government subsidy for FV $(c_{pf}^*, d_f^*)$ | $max\left(0, \frac{1}{\left(2+\frac{1}{\beta^2}\right)}\left[\left(G+\gamma_f+\frac{2c_0D}{v^2}\right)\left(1+\frac{1}{\beta^2}\right)+\frac{1}{2}c_c\left((1-\xi)L_T\right)^2+C_{AD}\beta^2v^2+\left(\frac{c_{d'}}{\beta v}\right)\right]\right)$ | $max\left(0, min\left(D, \frac{v^2}{2c_0}\left(G+\gamma_f-c_{pf}^*\right)+D\right)\right)$ |
| Considering government subsidy for PSP $(c_{pl}^*, d_l^*)$ | $max\left(0, \frac{1}{\left(2+\frac{1}{\beta^2}\right)}\left[\left(G+\frac{2c_0D}{v^2}\right)\left(1+\frac{1}{\beta^2}\right)+\frac{1}{2}c_c\left((1-\xi)L_T\right)^2+C_{AD}\beta^2v^2+\left(\frac{c_{d'}}{\beta v}\right)-\gamma_l\right]\right)$ | $max\left(0, min\left(D, \frac{v^2}{2c_0}\left(G+\gamma_f-c_{pl}^*\right)+D\right)\right)$ |
| Considering government subsidy for FV and PSP $(c_{plf}^*, d_{lf}^*)$ | $max\left(0, \frac{1}{\left(2+\frac{1}{\beta^2}\right)}\left[\left(G+\gamma_f+\frac{2c_0D}{v^2}\right)\left(1+\frac{1}{\beta^2}\right)+\frac{1}{2}c_c\left((1-\xi)L_T\right)^2+C_{AD}\beta^2v^2+\left(\frac{c_{d'}}{\beta v}\right)-\gamma_l\right]\right)$ | $max\left(0, min\left(D, \frac{v^2}{2c_0}\left(G+\gamma_f-c_{plf}^*\right)+D\right)\right)$ |

## 5. Numerical illustration

This section analyzes the best response strategies for the PSP and FV, specifically considering scenarios where the PSP addresses one service request at a time. Also, we show the sensitivity analysis on platoon velocity, cybersecurity investment, and malicious vehicle probability.

### *5.1 Numerical illustration for the PlaaS framework*

To check our proposed methodology's applicability and reliability, we first implement the Stackelberg game between the PSP and FV in connection with the PlaaS framework. The parameters are obtained from the literature and truck platooning reports and are mentioned in Table 4.

**Table 4.**
Description of the parameters and corresponding values (₹ - Indian Rupees)

| **Notation** | **Values** | **Notation** | **Values** |
|---|---|---|---|
| $c_d$ | ₹ 150 | $\xi$ | 0.5 |
| $c_f$ | ₹ 105 | $\psi$ | 0.25 |
| $c_o$ | ₹ 180 | $\eta$ | 0.5 |
| $c_c$ | ₹ 400 | $C_{dv}$ | 0.6 |
| $\gamma_F, \gamma_l$ | ₹ 50 | $C_{dp}$ | 0.42 |
| $v_p$ | 42 km/hr | $v$ | 60 km/hr |
| $D$ | 500 km | $A$ | 8 m$^2$ |
| $\rho_{diesel}$ | 850 kg/m$^3$ | $\rho_{air}$ | 1.225 kg/m$^3$ |

Based on the above parameters, we solved problems (1) and (2) to get the optimal distance traveled with platoon and service fee by PSP. The optimal distance to be travelled with a platoon is $411.10\ km,$ and the service fee charged by PSP is ₹ $206.07$. Based on this value, the total travelling cost for FV if it travels alone (i.e $d^* = 0$) for a total $D$ distance is ₹ $104544.12$. and if it travels $d^*$ with platoon then ₹ $96093.73$. Also, the total profit gained by PSP is ₹ $56301.18$.

### *5.2 $CO_2$ emission*

CAV Platooning significantly reduces aerodynamic drag force for FVs, leading to a quantifiable decrease in fuel consumption observed in different platoon projects. This reduction in fuel usage consequently results in lower carbon dioxide ($CO_2$) emissions. Assuming the FVs are diesel-powered trucks, the $CO_2$ emission savings can be quantified using the emission factor provided by the U.S. Environmental Protection Agency, which states that the combustion of one litre of diesel fuel emits approximately 2.69 kg of $CO_2$ [42], [43]. By applying this $CO_2$ emission factor ($\phi = 2.69\ kg$) in Eq. 8, along with the parameter values specified in Table 4, the total reduction in $CO_2$ emissions calculated per trip for a follower vehicle is approximately 47.04 kg.

### *5.3 Sensitivity analysis*

Based on the parameters mentioned in Table 4, we solved problems (1) and (2) and performed a sensitivity analysis varying the velocity reduction ratio $\beta$. Fig. 5 illustrates the sensitivity of the optimal follower vehicle distance ($d^*$) and the optimal service charge per unit distance ($c_p^*$) with changes in $\beta$. The parameter $\beta$ is defined as the ratio of the platoon velocity to the individual vehicle velocity $\left(\beta = \frac{v_p}{v}\right)$. The influence of $\beta$ helps evaluate the relative efficiency gained by traveling within a platoon, informing the decision-making of the platoon service provider (PSP) and follower vehicles (FVs).

Fig. 5 (a) shows that as $\beta$ increases from 0 to approximately 1.8, $d^*$ increases sharply, indicating that follower vehicles are incentivized to travel longer distances with the platoon when the speed is moderately lower than individual travel speeds. Beyond $\beta \approx 1.8$, $d^*$ approaches a minimal gain in follower participation with the platoon. Fig. 5 (b) shows that $c_p^*$ decreases sharply with increasing $\beta$, indicating that the PSP must reduce the service fee per km to encourage FV participation when the platoon speed is close to or exceeds that of individual vehicles. This analysis highlights a critical range of $\beta \leq 1.8$, where the PSP and FVs achieve optimal objective values. Moderate to high platoon speed advantages lead to greater follower participation, lower total travel costs for FVs, and higher profitability for the PSP as travel distance by FVs increases.

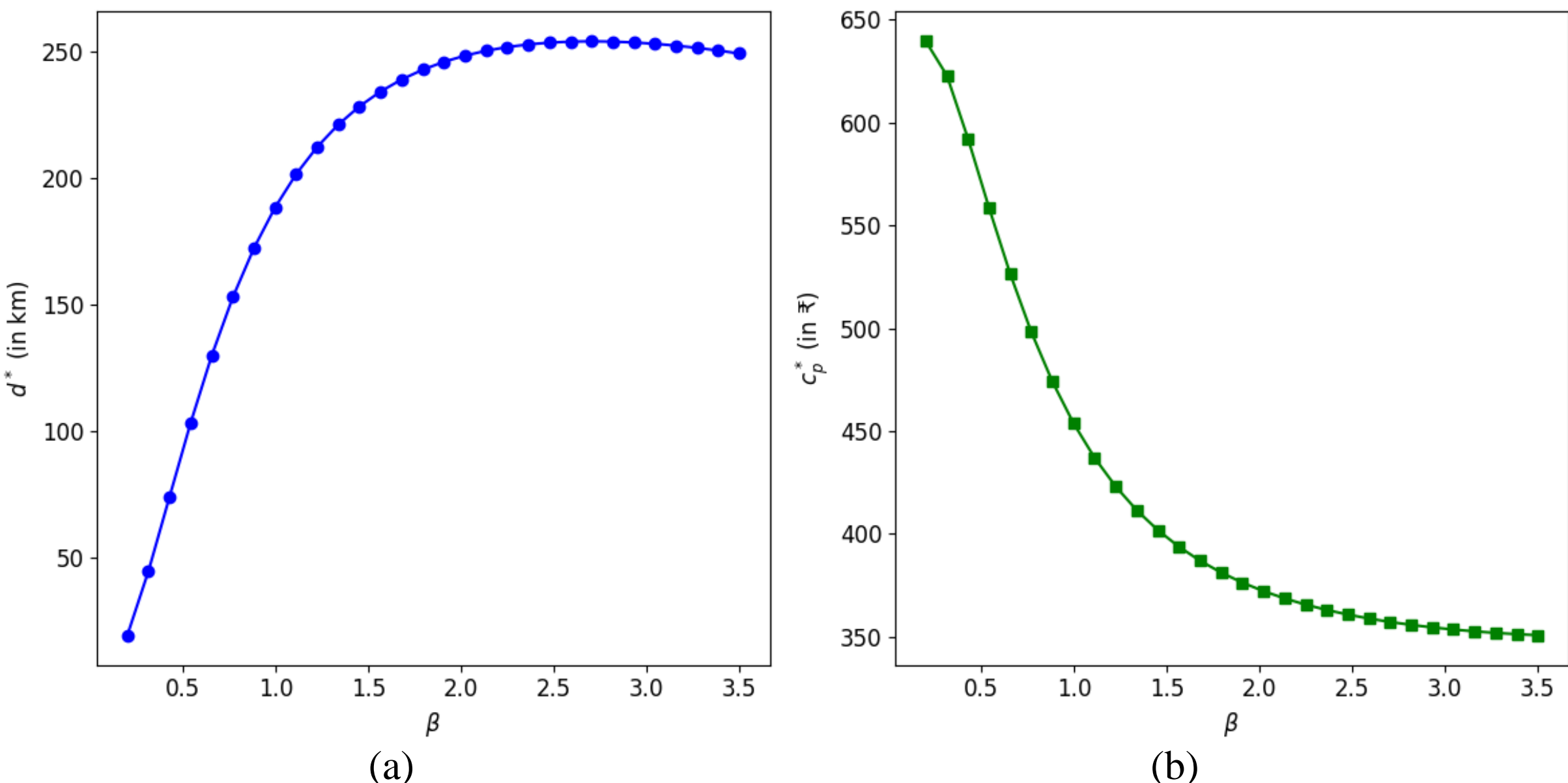

(a) (b)

**Fig 5.** The effect of velocity reduction ratio ($\beta$) (a) on optimal travel distance ($d^*$), (b) Optimal service charge ($c_p^*$)

Fig. 6 illustrates the effect of the velocity reduction ratio ($\beta$) and the ratio of the coefficient of air drag ($\alpha = c_{dp}/c_{df}$) between platooning ($c_{dp}$) and individual follower vehicles ($c_{df}$) on the optimal service charge per unit distance ($c_p^*$) and the optimal follower vehicle distance ($d^*$). In Fig. 6 (a), $c_p^*$ increases nonlinearly with both $\beta$ and $\alpha$. When $\beta$ is low, indicating that the platoon speed is significantly lower than that of an individual FV, the PSP sets a minimal $c_p^*$ to encourage participation. The service charge increases with $\beta$ until it reaches a peak, after which it declines. This pattern suggests that when the platoon velocity exceeds a certain threshold, FV participation decreases, prompting the PSP to reduce $c_p^*$, which can reduce overall platoon profitability. Higher $\alpha$ values represent greater aerodynamic drag reduction, which allows the PSP to charge higher service fees, as they contribute to fuel savings of FVs. This indicates that the service provider can set higher prices when aerodynamic benefits are substantial, and the platoon velocity is relatively low compared to the individual FV velocity.
Fig. 6 (b) shows that $d^*$ increases gradually with both $\beta$ and $\alpha$. As $\beta$ increases, follower vehicles are incentivized to travel longer distances with the platoon due to lower total travel costs from higher speeds. Similarly, greater values of $\alpha$ enhance fuel efficiency, encouraging longer participation of potential FVs. The observed smooth and increasing trend of $d^*$ with respect to both parameters highlight that improved fuel efficiency and minimal speed compromise make platooning more attractive for follower vehicles.

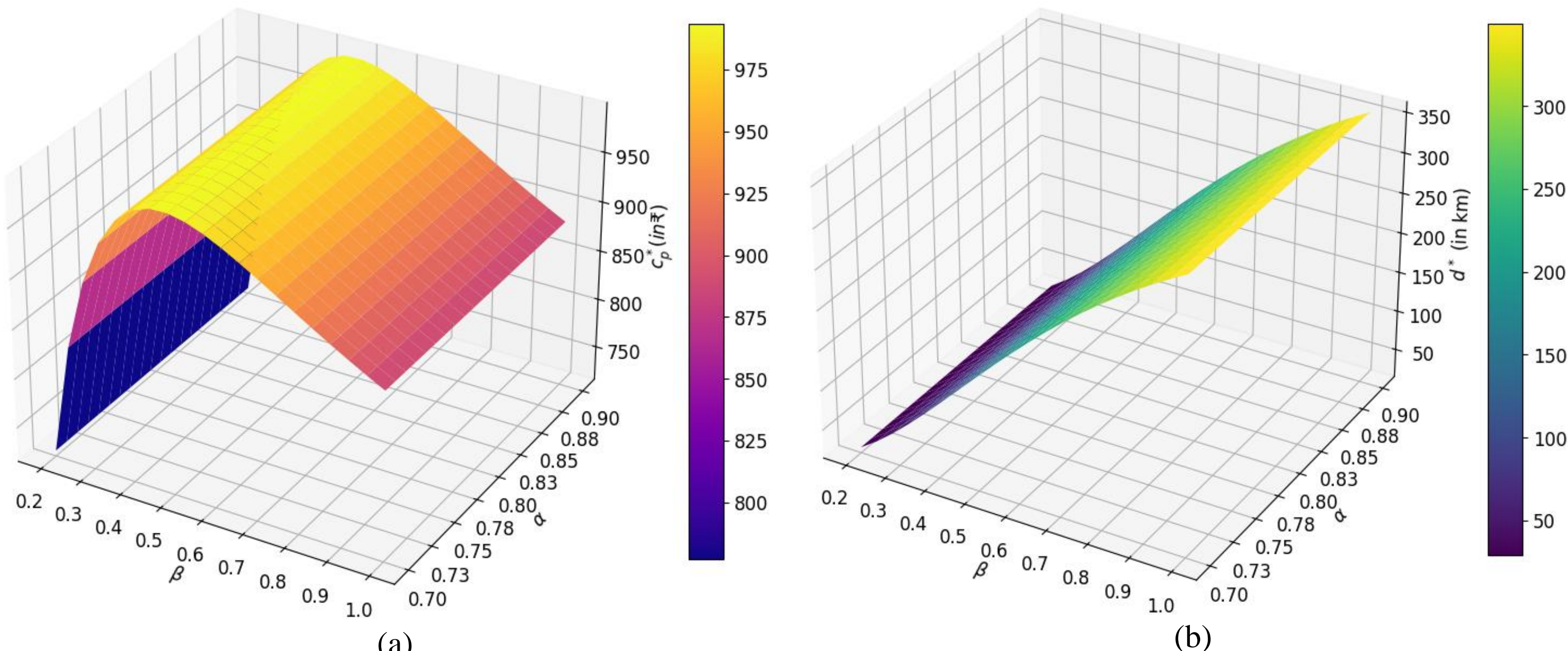

(a) (b)

**Fig 6.** Surface plot showing the effect of $\alpha$ and $\beta$ on (a) the optimal value of $c_p^*$ (b) optimal value of $d^*$

## 6. Managerial Implication

In a CAV platoon, the travel time for FVs directly affects both the total travel cost and the PSP's profit. Hence, vehicles with higher delay cost may travel shorter distances with the platoon, as minimizing travel time is critical. Fig. 7 examines how the choice of platoon velocity influences total travel cost and PSP profit under different follower vehicle (FV) delay costs ($c_d$), ranging from $c_d = 0$ (no urgency) to $c_d = 150$ (significant urgency).

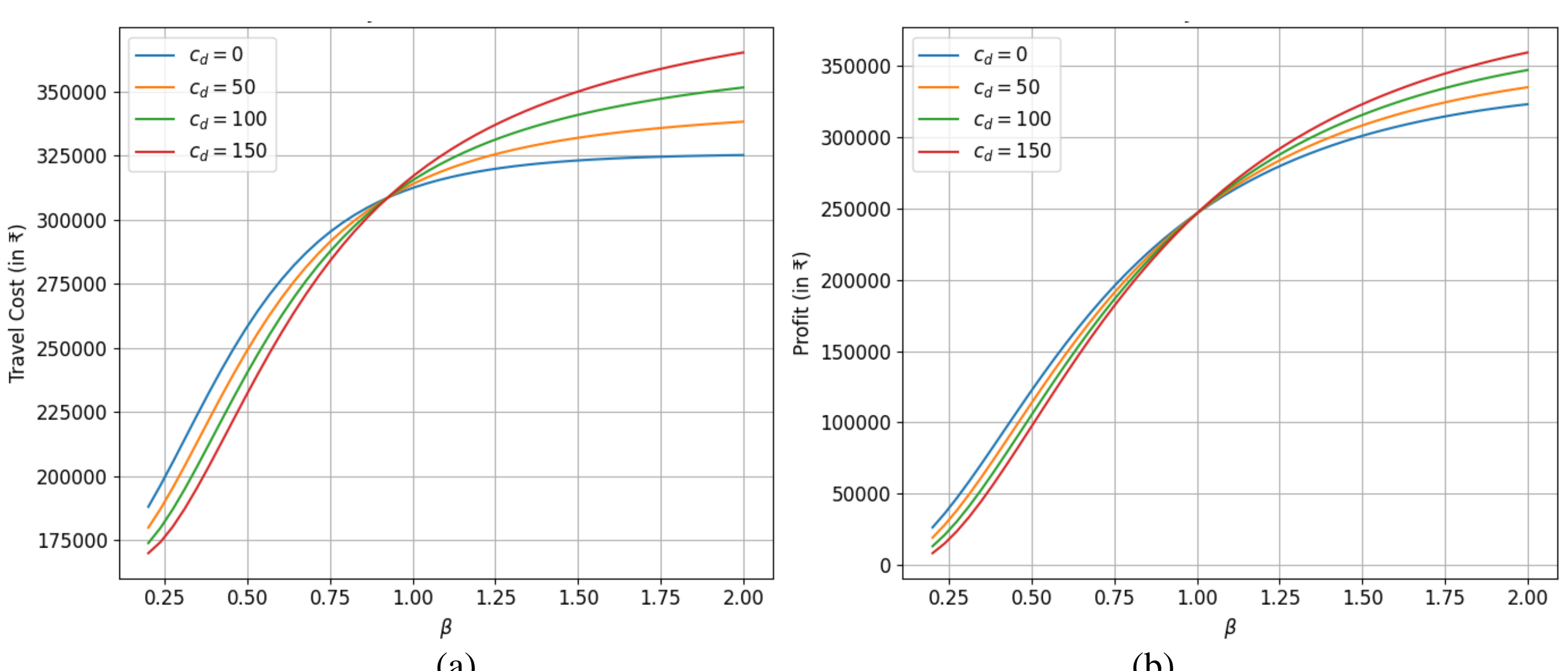

(a) (b)

**Fig 7.** Effect of $\beta$ and follower delay cost ($c_d$) on (a) total travel cost of the follower and (b) service provider profit

In Fig. 7 (a), total travel cost increases with $\beta$, with steeper rises observed at higher $c_d$ values. For $\beta < 1$, which indicates the average platoon velocity $(v_p)$ is lower than that of the vehicle traveling independently $(v)$, vehicles with higher delay costs travel shorter distances with the platoon, leading to lower travel costs and reduced PSP profit compared to low-delay-cost vehicles. In Fig. 7 (b), PSP profit grows with $\beta$ up to a certain point, after which it levels off. Higher $c_d$ values consistently lead to greater profits at higher $\beta$ values ($\beta > 1$), as high delay-cost vehicles are willing to pay more to minimize travel time.

Therefore, PSP gains a higher profit from high delay-cost vehicles performing time-critical operations with higher platoon velocity. However, such a situation may not arise in real life, as platoons generally don't travel faster than independent vehicles. In contrast, the PSP's profit is higher for low delay-cost vehicles when traveling at a slower platoon velocity ($\beta < 1$).

Fig. 8 shows that the reduction in $CO_2$ emission depends on the $\beta$ and the total government subsidy ($\gamma = \gamma_l + \gamma_f$) provided to the PSP and FVs per kilometre of travel. Higher subsidies reduce $CO_2$ emission significantly, peaking at $\beta \approx 0.8$; beyond this, greater speed differences increase fuel consumption as given in Eq. 8. Without subsidies ($i.e. \gamma = 0$), the model predicts negligible emission reductions. These results highlight that targeted financial incentives are essential for encouraging efficient platooning and achieving significant environmental benefits.

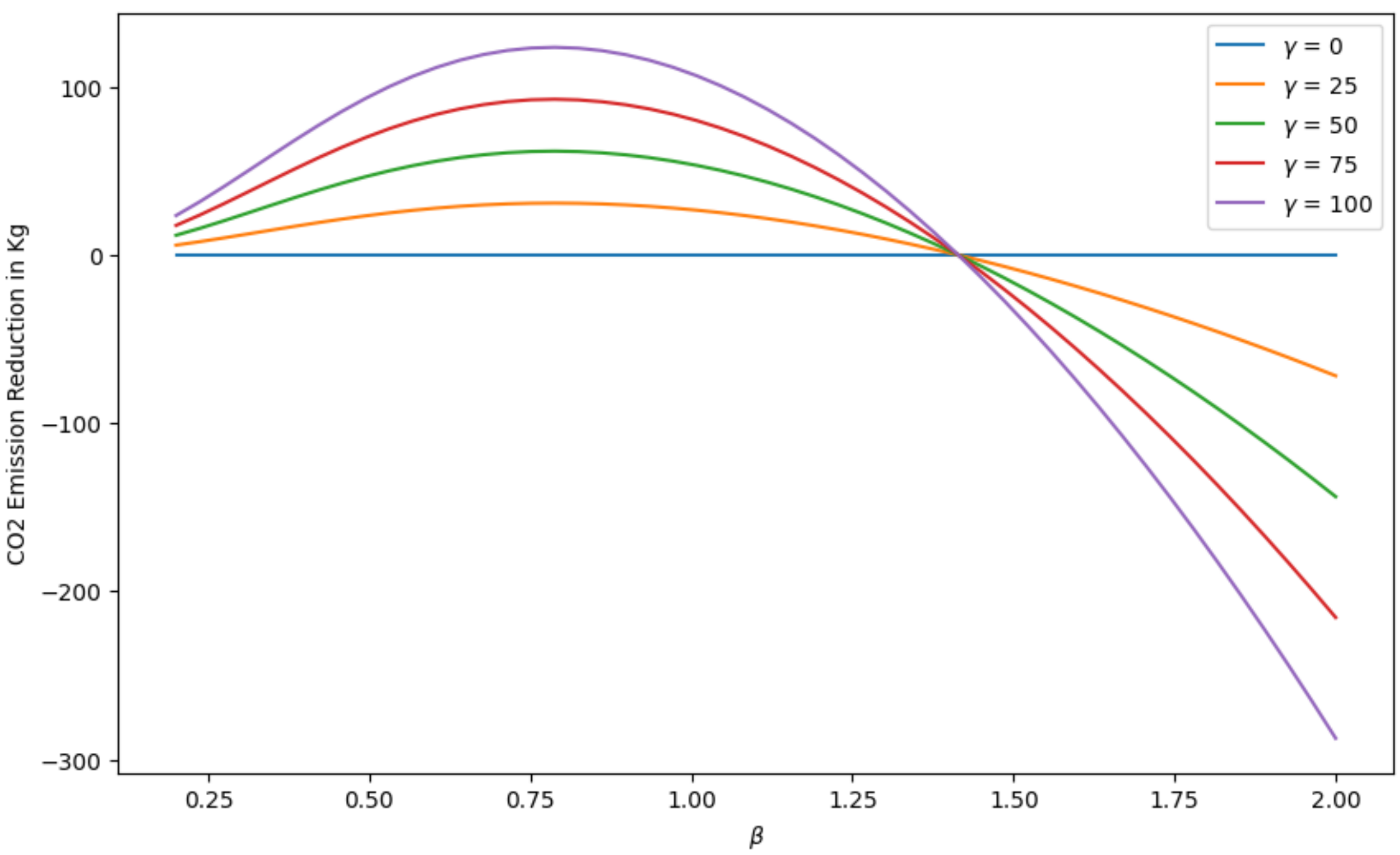

**Fig 8.** Effect of government subsidy $(\gamma)$ and velocity reduction ratio $(\beta)$ on reduction in $CO_2$ emission using CAV Platooning

Fig. 9 shows how the value of $c_p^*$ and $d^*$ changes with the $\beta$ across different delay cost levels $(c_d)$. The relationship between $\beta$, $c_d$, and $c_p^*$ is nonlinear as observed in Fig. 9. When $\beta < 1$, indicating that the platoon is traveling slower than the individual FV speed, vehicles with higher urgency achieve the same travel distance at a lower optimal fee compared to less urgent vehicles. As $\beta$ approaches 1, where platoon velocity matches the original vehicle velocity at that point both $c_p^*$ and $d^*$ converge for all $c_d$ values, and total profit remains constant across all vehicle types. This convergence point around $\beta \approx 1$ implies that operational speed constraints and economic delay penalties offset each other, offering valuable guidance for designing pricing strategies and service structures in platoon systems.

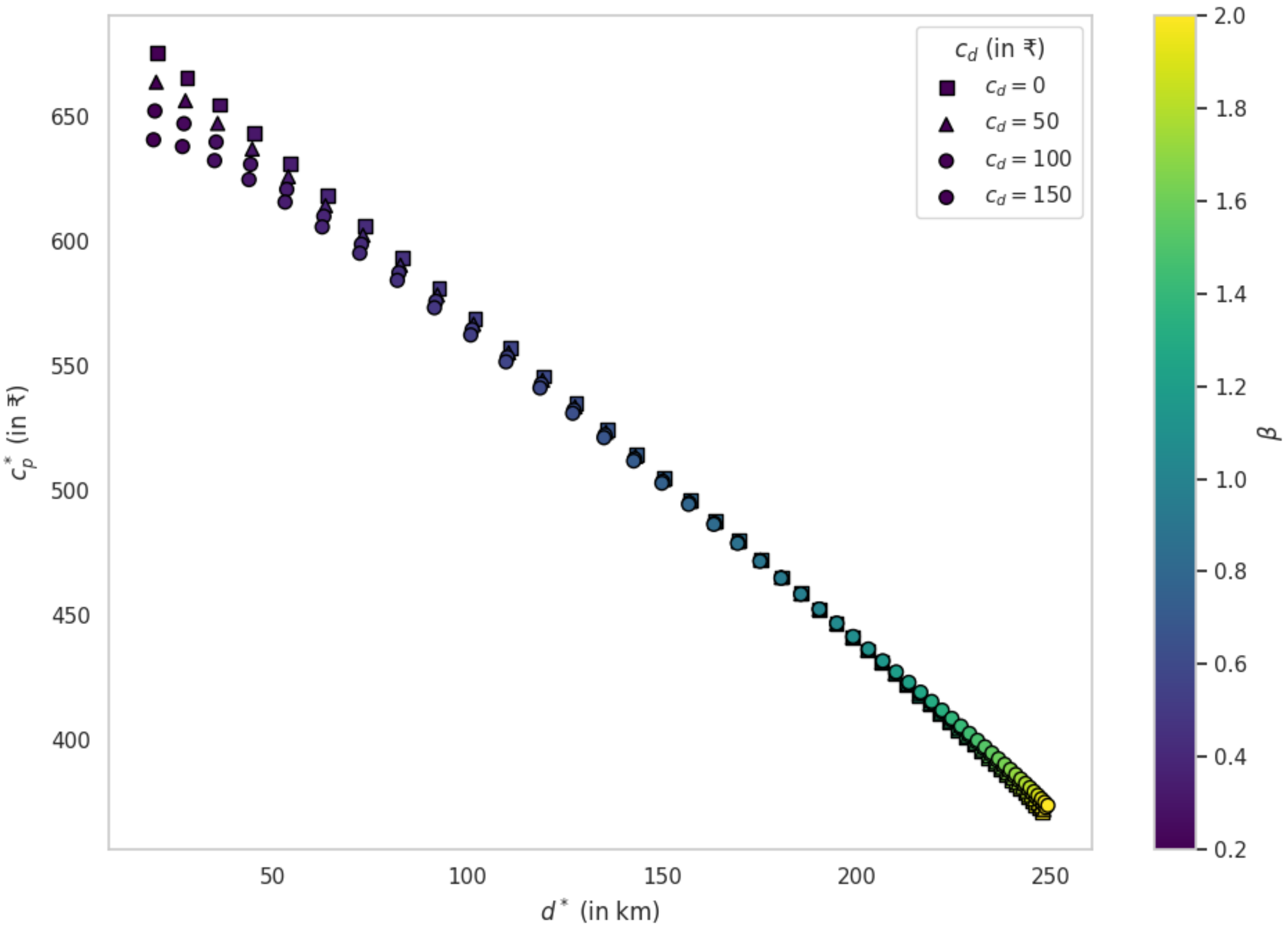

**Fig 9.** Scatter plot of optimal distance $d^*$ vs. service charge $c_p^*$, colored by velocity reduction ration $\beta$ and marked by FVs delay cost $c_d$.

## 7. Conclusion and Future Work

In this study, we designed the Platooning as a Service (PlaaS) framework for sustainable transportation. We present a cost-benefit business model, where the platoon service provider (PSP) sets an optimal service fee in exchange of the benefits offered to the potential follower vehicle (FV), and the FV decides the optimal distance to be travelled with the platoon to minimize its travel costs. We considered different operational costs such as fuel, delay, cognitive and computational loads, and government subsidy policies, that help model identify optimal strategies for both PSP and FVs. This decision problem has been modelled using a Stackelberg game and solved using the KKT conditions. The models' effectiveness and decision parameter threshold are demonstrated with a simulated example and sensitivity analysis. Analytical solution validates that the PSP's optimal service fee and the FV's optimal travel distance with platoon are highly sensitive to velocity reduction and delay costs. Our analysis also shows that government subsidies to both players incentivize platoon formation, which significantly reduces carbon emissions. Thus, the framework offers a comprehensive solution for designing cost-effective and sustainable CAV-based platooning systems.

Several limitations in the current PlaaS framework that may be addressed in future extensions of this research are listed below.

1) We have not considered the uncertainty of demand for platooning services in the revenue model. The effect of demand elasticity as a function of price over a period and traffic congestion should be explored.

2) Our model does not consider the exact position of the FV joining the platoon. The benefits (fuel consumption) and platoon dynamics may vary based on the position in the platoon. The platoon size may also impact the stability and road utilization efficiency. These aspects need to be studied in the future.
3) Our study only assumes a homogeneous platoon, where all the vehicles are assumed to have similar characteristics in terms of fuel savings. A natural extension to this work could be exploring the impact of heterogeneous platoons.
4) The proposed model considers the single FV and service provider problem. These can be further improved by considering the competition between the multiple service providers, interaction among multiple FVs, and their price selection to form the platoon.
5) Cybersecurity solutions for CAVs must be developed leveraging the recent breakthroughs in federated learning and edge security to ensure secure platooning operations.

In addition, the distribution of computational resources and the effect of switching between manual and autonomous driving may be studied in more detail. Furthermore, techniques such as reinforcement learning may be implemented to derive optimal policies for platooning operations.

## Appendix A

### A.1.1 Proof of Proposition 1.1

Proposition 1.1 is to prove that the follower vehicle's objective problem is a convex problem. To prove this, we show that the objective function is convex, and the constraints are also convex by showing $\frac{\partial c_{FV}}{\partial d} \leq 0$ and $\frac{\partial^2 c_{FV}}{\partial d^2} \geq 0$. Where, $c_{FV} = c_{ind} + c_{pla}$, the follower vehicle's objective function. The partial differentiation is given as below:
The function given in Section 3.2 is quadratic with single decision variable $d$. Therefore, as given in [44] the optimization problem is convex is if the objective function is convex for minimization and all the constraint are convex. The convexity of objective function with single variable is proved by taking partial differentiation:

$$\frac{\partial^2 c_{FV}}{\partial d^2} = \frac{2c_0}{v^2} \geq 0$$

Thus $c_{FV}$ is a convex function w.r.t $d$.

### A.1.2 Proof of Proposition 1.2

For the follower vehicle's optimization problem, the constraint $d \leq D$ is a linear form function, then the feasible set is also convex. Thus, we prove follower vehicle's optimization problem is a convex problem. Since, there is non-negativity constraint related to decision variable $d$, to solve nonlinear optimization problem KKT conditions are use. the non-negative constraints take the form $-d \leq 0$.
From the Proposition 1.1 the objective function $c_{FV}$ is also continuous and twice differentiable. Thus, the Lagrangian form of follower vehicle's optimization problem (2) can be written as below:

$$L(d,\ \lambda_1, \lambda_2) = \frac{(D-d)}{v} c_d + TC_{df} v^2 (D-d) c_f + c_o \frac{(D-d)^2}{v^2} + \left(\frac{d}{v_p}\right) c_d + TC_{dp} v_p^2 d c_f + \frac{1}{2} c_c d (\xi L_T)^2 + d c_p - \gamma_f d - \lambda_1 (d) + \lambda_2 (d - D)$$

According to the necessary condition given by KKT, if $\boldsymbol{d}^*$ is a local minimum, then there exist KKT multipliers $\lambda_1$ and $\lambda_2$ such that the following optimality conditions are satisfied:

1) Stationary condition:

$$\frac{\partial L}{\partial d_i} = -\frac{2c_o(D-d)}{v^2} - C_{df}Tc_f v^2 + C_{dp}Tc_f v_p^2 + \frac{c_d}{v_p} - \frac{c_d}{v} + c_p + \frac{1}{2}c_c(\xi \times L_T)^2 - \gamma_F - \lambda_1 + \lambda_2$$

2) Primal feasibility:

$$d - D \leq 0$$
$$-d \leq 0$$

3) Complementary slackness:

$$\lambda_2(d - D) = 0$$
$$\lambda_1(-d) = 0$$

4) Dual feasibility:

$$\lambda_1,\ \lambda_2 \geq 0$$

$d^* = 0$ and $d^* = D$ are the two cases in which the constraints are binding, and they cannot occur together. This implies that $\lambda_1^*$ and $\lambda_2^*$ are linearly independent when the corresponding constraints are binding. Hence $\lambda_1^*$ and $\lambda_2^*$ are regular points.
From Proposition 1.1 and Proposition 1.2, we conclude that the follower vehicle's problem is convex optimization problem, which means that KKT conditions are sufficient and if it satisfied then $\boldsymbol{d}^*$ is a global optimal solution.

**A.2.1 Proof of Proposition 2.1**

Proposition 2.1 is to prove that the PSP objective problem is a convex problem. To prove this, we show that the objective function is concave (for maximization), and the constraints are convex by showing $\frac{\partial w_{PSP}}{\partial c_p} \geq 0$ and $\frac{\partial^2 w_{PSP}}{\partial c_p^2} \leq 0$. The partial differentiation is given as below:
The function given in Section 3.3 is quadratic with single decision variable $c_p$ after inserting $d^*$ in $w_{PSP}$. Therefore, as given in [44] the optimization problem is convex is if the objective function is concave for maximization and all the constraint are convex. The convexity of objective function with single variable is proved by taking partial differentiation:

$$\frac{\partial^2 w_{PSP}}{\partial c_p^2} \leq 0$$

Thus $w_{SP}$ is a convex function w.r.t $d$.

**A.2.2 Proof of Proposition 2.2**

For the PSP optimization problem, the constraint $c_p \geq 0$ is a linear form function, then the feasible set is also convex. Thus, we prove PSP optimization problem is a convex problem. Since, there is non-negativity constraint related to decision variable $c_p$, to solve nonlinear optimization problem KKT conditions are use. the non-negative constraints take the form $-c_p \leq 0$.
From the Proposition 2.1 the objective function $w_{PSP}$ is also continuous and twice differentiable. Thus, the Lagrangian form of follower vehicle's optimization problem (2) can be written as below:

$$L(d,\ \lambda_1, \lambda_2) = c_p d^* + \gamma_l d^* - c_{d'}\left(\frac{d^*}{v_p}\right) - \frac{1}{2}c_c\big((1-\xi)L_T\big)^2 d^* - c_o\left(\frac{d_l^2}{v_p^2}\right) - TC_{dl}v_p^2 d_l c_f - \theta_1(c_p)$$

According to the necessary condition given by KKT, if $\boldsymbol{c_p^*}$ is a local minimum, then there exist KKT multipliers $\theta_1$ such that the following optimality conditions are satisfied:

1) Stationary condition:

$$\frac{\partial L}{\partial c_p} = -\frac{v^2 c_p}{2c_0} + \left(\frac{v^2}{2c_0}\right)\left[-c_p - \frac{1}{2}c_c(\xi L_T)^2 + \gamma_f + TC_{df}v_p^2 c_f + c_d\left(1-\frac{1}{\beta}\right)\right]$$
$$+\frac{v^2}{v_p^2}\left(\left(\frac{v^2}{2c_0}\right)\left[-c_p - \frac{1}{2}c_c(\xi L_T)^2 + \gamma_f + TC_{df}v_p^2 c_f + c_d\left(1-\frac{1}{\beta}\right)\right] + D\right)$$
$$+\frac{1}{4c_0}c_c\big((1-\xi)L_T\big)^2 - \frac{v^2\gamma_l}{2c_0} + \frac{TC_{df}v_p^2 c_f v^2}{2c_0} + \frac{c_{d'}v^2}{2c_0 v_p} + D = 0$$

2) Primal feasibility:

$$-c_p \leq 0$$

3) Complementary slackness:

$$\theta_1(-c_p) = 0$$

4) Dual feasibility:

$$\theta_1 \geq 0$$

From Proposition 2.1 and Proposition 2.2, we conclude that the PSP problem is convex optimization problem, which means that KKT conditions are sufficient and if it satisfied then $\boldsymbol{c_p^*}$ is a global optimal solution.

**A. 3**. **Proof of Proposition 3**

Let $d^*_{with\ subsidy}$ and $d^*_{w/o\ subsidy}$ be the optimal travel distances of a follower vehicle under subsidized and unsubsidized conditions, respectively as given from Eq. 7.

$$d^*_{with\ subsidy} = \frac{v^2}{2c_0}\left(G + \gamma_f - \frac{1}{\left(2+\frac{1}{\beta^2}\right)}\left[\left(G + \gamma_f + \frac{2c_0 D}{v^2}\right)\left(1+\frac{1}{\beta^2}\right) + \frac{1}{2}c_c\big((1-\xi)L_T\big)^2 + C_{AD}\beta^2 v^2 + \left(\frac{c_{d'}}{\beta v}\right) - \gamma_l\right]\right) + D$$

$$d^*_{w/o\ subsidy} = \frac{v^2}{2c_0}\left(G - \frac{1}{\left(2+\frac{1}{\beta^2}\right)}\left[\left(G + \frac{2c_0 D}{v^2}\right)\left(1+\frac{1}{\beta^2}\right) + \frac{1}{2}c_c\big((1-\xi)L_T\big)^2 + C_{AD}\beta^2 v^2 + \left(\frac{c_{d'}}{\beta v}\right)\right]\right) + D$$

The net gain in distance due to subsidies is:

$$\Delta d = d^*_{with\ subsidy} - d^*_{w/o\ subsidy}$$
$$= \frac{v^2}{2c_0} \times \frac{(\gamma_l + \gamma_f)}{\left(2+\frac{1}{\beta^2}\right)}$$

$$= \frac{v^2}{2c_0} \times \frac{\beta^2(\gamma_l + \gamma_f)}{(2\beta^2 + 1)}$$

The fuel consumption when FV need to travel total $D$ distance is given in Table 2 as:

$$Fuel\ consumption = TC_{df}v^2(D - d^*) + T\alpha C_{df}\beta^2 v^2 d^*$$

Therefore, the total fuel saved due to the government subsidy in one trip is:

$$\Delta Fuel = Fuel\ consumption\ without\ subsidy - Fuel\ consumption\ with\ subsidy$$

$$= TC_{df}v^2\left(D - d^*_{w/osubsidy}\right) + T\alpha C_{df}\beta^2 v^2 d^*_{w/osubsidy} - TC_{df}v^2\left(D - d^*_{with\ subsidy}\right) - T\alpha C_{df}\beta^2 v^2 d^*_{with\ subsidy}$$

$$\Delta Fuel\ (litre) = TC_{df}v^2 \times \frac{v^2}{2c_0} \times \frac{\beta^2(\gamma_l + \gamma_f)}{(2\beta^2 + 1)}(1 - \alpha\beta^2)$$

Since 1 litre of fuel burnt approximately emits $\phi$ kg of $CO_2$, the saving of $CO_2$ emission in the one trip equals:

$$\Delta CO_2(kg) = T\alpha\phi C_{df}v^2 \times \frac{v^2}{2c_0} \times \frac{\beta^2(\gamma_l + \gamma_f)}{(2\beta^2 + 1)}(1 - \alpha\beta^2)$$

This proposition confirms that government subsidies incentivize optimal platooning behaviour and produce quantifiable environmental benefits by reducing $CO_2$ emissions. From a sustainability and policy-making perspective, this result supports the integration of monetary incentives in green transportation planning.